\documentclass[review]{elsarticle}

\usepackage{hyperref}

\journal{Nuclear Instruments and Methods in Physics Research A}

\usepackage{graphicx}

\begin{document}

\begin{frontmatter}

%Title of paper
\title{Sensitive neutron detection method using delayed coincidence transitions in existing iodine-containing detectors}

\author[jinr-adr]{E. Yakushev\corref{mycorrespondingauthor}}
\cortext[mycorrespondingauthor]{Corresponding author}
\ead{yakushev@jinr.ru}

\author[jinr-adr]{S. Rozov}

\author[mit-adr]{A. Drokhlyansky}

\author[jinr-adr]{D. Filosofov}

\author[jinr-adr,cu-adr]{Z. Kalaninova}

\author[jinr-adr]{V. Timkin}

\author[jinr-adr]{D. Ponomarev}

\address[jinr-adr]{Dzhelepov Laboratory of Nuclear Problems, JINR, Joliot-Curie 6, 141980 Dubna, Russia}
\address[mit-adr]{Department of Physics, MIT, Cambridge, MA 02139, USA}
\address[cu-adr]{Department of Nuclear Physics and Biophysics, Comenius University, 84248 Bratislava, Slovakia}

%\date{\today}

\begin{abstract}
This work explains a new, highly sensitive method for the detection of neutrons,
which uses the T$_{1/2}=845$~ns delay in the decay of $^{128}$I
at the 137.8~keV energy level, resulting from the capture of thermal neutrons by iodine nuclei in NaI and CsI
scintillation detectors. The use of delayed coincidence techniques with a several $\mu {\rm s}$ delay time window
for delayed events allows for the highly effective discrimination of neutron events from any existing background signals.
A comparison of ambient neutron measurements between those identified through the suggested method from
a cylindrical, \o$\, 63 \, {\rm mm}\times 63\, {\rm mm}$ NaI(Tl) scintillator and those from a low-background
proportional $^3$He counter experimentally demonstrates the efficacy of this neutron detection method.
For an isotropic, $4\pi$, thermal neutron flux of 1~${\rm n}\, {\rm cm}^{-2}\, {\rm s}^{-1}$, 
the absolute sensitivity of the NaI detector was found to be $6.5 \pm 1\, {\rm counts}\, {\rm s}^{-1}$ 
with an accidental coincidence background of $0.8\, {\rm events}\, {\rm day}^{-1}$ for any delay time window of $\Delta {\rm t}=1\, \mu $s.
The proposed method can provide low-background experiments,
using NaI or CsI, with measurements of the rate and stability of incoming neutron flux to a greater accuracy
than 10$^{-8}\, {\rm n}\, {\rm cm}^{-2}\, {\rm s}^{-1}$.
\end{abstract}

\begin{keyword}
Neutron detection\sep NaI \sep CsI
%\MSC[2010] 00-01\sep  99-00
\end{keyword}

\end{frontmatter}

%\linenumbers

%\maketitle

\section{Introduction}

The search for alternative neutron detectors to replace those
that use $^3$He presents a pressing issue in modern
physics~\cite{Kouzes20101035, Kouzes2015172}. The fairly low cost, widespread availability,
and ease of use of NaI(Tl), CsI(Tl), CsI(Na), and CsI(pure) scintillation detectors makes their implementation
for the detection of neutrons an incredible opportunity.
Neutrons have previously been detected by these scintillation devices through various approaches (see Table~\ref{tab:nai_previous}).

\begin{table}[h]
\caption{Prior approaches for neutron detection with NaI and CsI\label{tab:nai_previous}}
\small\begin{tabular}{lllp{0.5\textwidth}}\hline
Reference & Detector & Neutrons & Short description  \\\hline
 \cite{Metwally201448}     &  NaI   & thermal & boron lining with available NaI detectors \\
 \cite{Holm201359}         & NaI & thermal & high-energy photons following (n,$\gamma$) reactions in the NaI \\
 \cite{refId0, Bernabei2008297} & NaI(Tl) 
& thermal & triple $\beta-\gamma-\gamma$ coincidences in two detectors following (n,$\gamma$) reactions on $^{23}$Na \\
 \cite{Kuznetsov2009} &   NaI & thermal & activated NaI detector ($^{128}$I $\beta-$decay, T$_{1/2}=25\, {\rm min}$ 
and $^{24}$Na $\beta-$decay, T$_{1/2}=15\, {\rm h}$)\\
 \cite{Collar201556}       &  CsI(Na) & fast & 57.6~keV signal from $^{127}$I(n,n$'$) inelastic scattering \\
 \cite{Bartle199954}       & NaI(Tl), CsI(Tl) & fast &  1-200~MeV neutrons, (n,p) and (n,$\alpha$) reactions, 
pulse-shape discrimination \\\hline
 \end{tabular}
\end{table}

It is important to note that, in~\cite{Holm201359}, when a NaI spectrometer was compared to a $^3$He-based portal
monitor with a comparable active volume, the detection efficiencies and minimum detectable activities of
the devices were found to be similar.

In general, neutron measurements with NaI and CsI detectors require accurately
accounting for background signals in the neutron detection region of interest because of the high sensitivity of
these detectors to cosmic and  $\gamma$-rays.
To efficiently discriminate neutron events from background signals, the neutron detection method proposed in this work uses delayed 
coincidence techniques with a 
short delay time window.

\section{Description of the method}

Iodine has only one stable isotope: $^{127}$I.
This isotope has a cross section for thermal neutron capture
of $\sigma_\gamma^{\rm 0}=6.2 \pm 0.2$~b~\cite{tableofisotopes},
which is 860 times lower than $^3$He's cross section of $5333 \pm 7$~b.
However, it is important to note that NaI (solid) has 109 times as many moles as an
equal volume of $^3$He (gas, 500 kPa), and CsI (solid) has 77 times as many moles.
Thus, despite its relatively low cross section, the fairly small \o$\, 63\, {\rm mm}\times63\, {\rm mm}$ NaI(Tl) detector 
(used for the experiment described in this work, and henceforth referred to with $^{63\times 63}$NaI) has an $\sim$50\% 
neutron capture (on $^{127}$I) effectiveness, as obtained from Geant4~\cite{Agostinelli2003250} modeling based on 
assumptions about the uniform thermal neutron flux (the Maxwell-Boltzmann distribution at room temperature).
$^{128}$I in its excited state (6.8~MeV) is produced as a result of neutron capture,
and its decay to the ground state proceeds through a series of low-energy levels
(see Table~\ref{tab:127i-scheme}). This decay process often includes the energy
level 137.8~keV with T$_{1/2}=845\pm 20$~ns.
To identify neutron capture by iodide in a detector, thereby measuring neutron flux, measurements of the following delayed coincidences 
by the same detector
can be used: the decay transition from 6.8~MeV to 137.8~keV (the neutron prompt signal) and the transition
from 137.8~keV to the ground state (the neutron delayed signal).

\begin{table}%
 \caption{$^{128}$I excited energy levels below 200~keV and their decay transitions (from ~\cite{tableofisotopes} and
\cite{SAKHAROV1991317}).
\label{tab:127i-scheme}}
\small\begin{tabular}{lll}\hline
Energy level  & T$_{1/2}$   & Energy levels\\
in $^{128}$I (keV) &  (ns) &  following decay (keV) \\\hline
180 && 160.8, 85.5, 27.4\\
167.4 &	175$\pm$15 &		137.8 \\
160.8	&&	27.4, 0  \\
151.6	&&	85.5, 27.4   \\
144.0 	&&	133.6   \\
{\bf 137.8}& {\bf 845$\pm$20} &	{\bf 133.6, 85.5} \\
133.6&	12.3$\pm$0.5	&	85.5, 27.4, 0  \\
128.2&&	85.5 \\
85.5&&	27.4 \\
27.4&&	0   \\
0&& \\\hline
 \end{tabular}
 \end{table}

According to~\cite{SAKHAROV1991317}, following neutron capture by iodide,
40$\pm$10\% of all de-excitations pass through the 137.8~keV energy level.
This energy level is filled and emptied via a series of low-energy, highly converted transitions,
including the 4.2~keV transition, which has never been experimentally investigated on its own.
These transitions raise uncertainty about the probability of decay via this energy level, causing uncertainty 
when estimating the detector's overall sensitivity.
The lack of direct measurements of low-energy $\gamma$-transition intensities and of internal conversion coefficients
also results in significant uncertainty in the detector's effectiveness at recording neutron prompt and delayed events.
Nevertheless, Geant4 MC's estimations obtain a $>$75\% absorption probability for low-energy $\gamma$-rays
and electrons released in the $^{63\times 63}$NaI detector, following neutron capture.
Thus, the overall effectiveness of this NaI detector at 
detecting neutrons is roughly estimated with performed MC to 
be $\sim$10\% and has significant uncertainties, as described above.  

\section{The Experiment}

\begin{figure}
\includegraphics[width=1.0\textwidth]{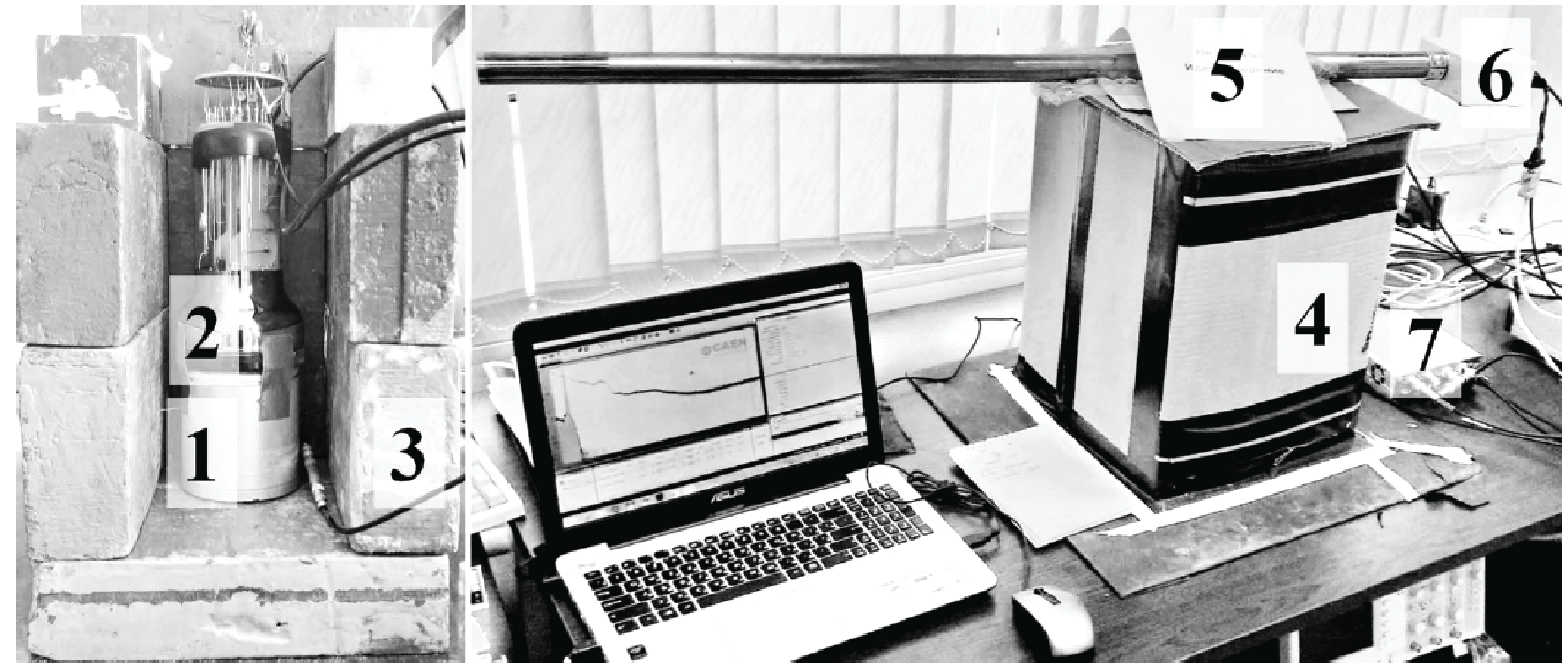}%
\caption{On left: NaI spectrometer inside lead shield
(front side of shield removed). 1) \o$\, 63\times 63$~mm NaI(Tl) scintillator,
2) R6091 PMT, 3) lead bricks.
On right: NaI and $^3$He neutron spectrometers during data acquisition.
4) Black box with NaI detector inside,
5) CHM-57 $^3$He filled proportional counter,
6) $^3$He neutron detector's preamplifier,
7) DT5780 Dual Digital Multi Channel Analyzer used to acquire
data from both detectors.\label{fig:nai-spectrometer}}
\end{figure}

For experimental verification of the suggested method, a simple spectrometer was created,
as shown in Fig.~\ref{fig:nai-spectrometer}.
A $3''$~PMT R6091 (Hamamatsu) was attached to a cylindrical, \o$\, 63\, {\rm mm}\times 63\, {\rm mm}$ NaI(Tl) scintillator via optical grease,
and this device was placed in a 5-cm thick lead brick well.
The top part of this lead brick shield was left open. A cardboard box, covered along the inside in black paper,
shielded the detector from light. A CAEN DT5780 Dual Digital Multi Channel Analyzer
collected data using an event by event list mode.
Consecutive events within a delay time smaller than $1.8\, \mu{\rm s}$
could not be properly registered due to the NaI scintillator's $\sim$250$\, {\rm ns}$ decay time,
after-pulses in the PMT, and impulse shapes with tails extending up to
approximately $1.5\, \mu {\rm s}$ after the initial peak.
Therefore, $1.8\, \mu{\rm s}$ was chosen as the trigger holdoff.
With an energy threshold slightly below 100~keV, the total count rate of the detector was $\sim$7.3~Hz.

\begin{figure}
 \includegraphics[width=1.0\textwidth]{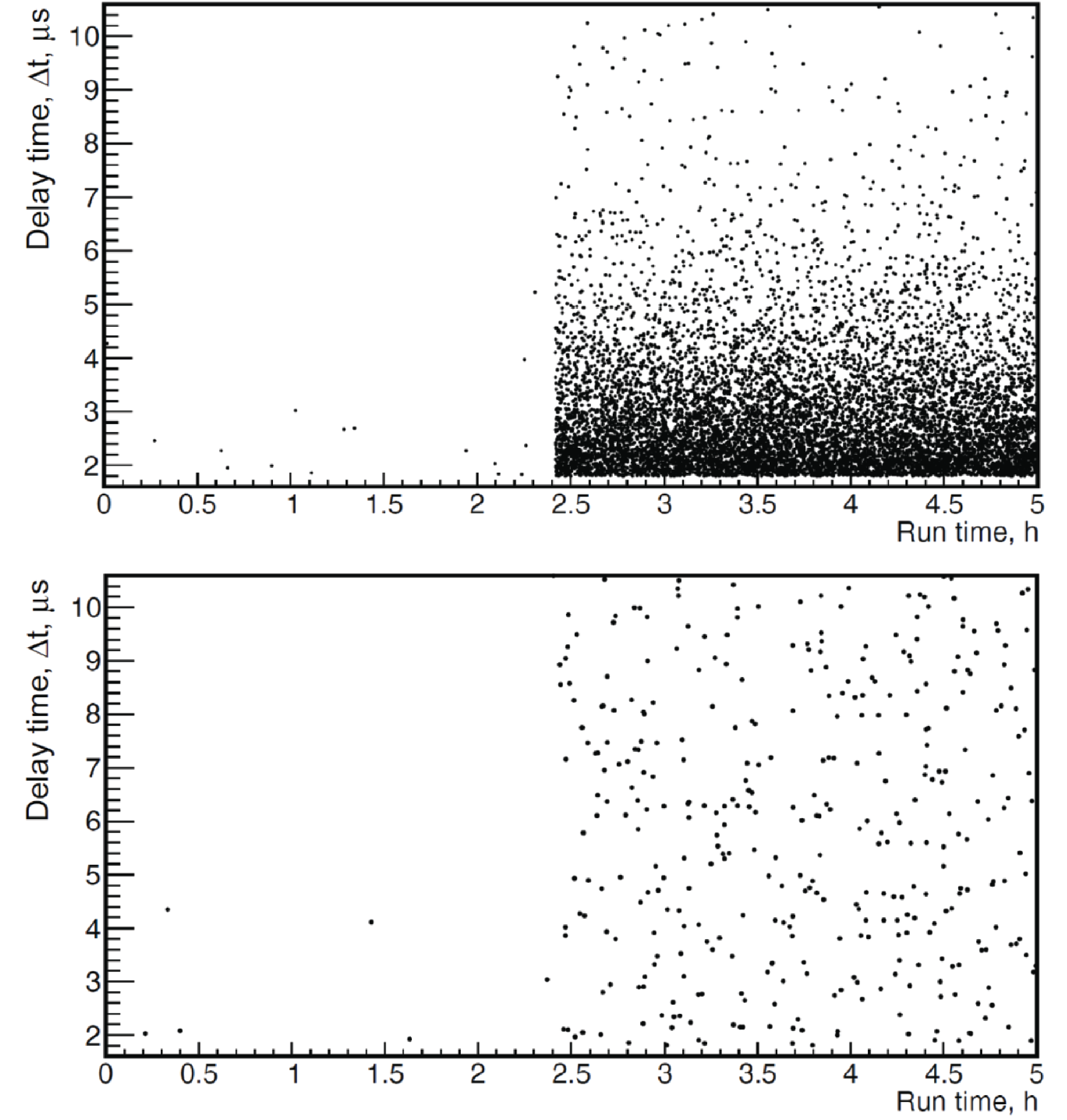}%
 \caption{Delayed coincidences registered without (left half of each graph) and in the presence of (right 
half of each graph) the PuBe neutron source. Each dot represents one coincidence event. 
The upper graph shows coincidences with delayed events in the 137~keV region, i.e. 
ADC channels from 150 to 350 (see Fig.~\ref{fig:nai-spectra}). During the run with the PuBe source, the number 
of coincidence events in the upper graph is evidently greater for a delay time of 
$\Delta$t$<$7~$\mu$s than for a larger delay time. The lower graph shows delayed coincidence 
events registered from ADC channel 1000 to 8000 
(well above the expected signal for a neutron event). In this graph, the noticeable increase of random coincidences  
for the run with the PuBe source, as compared to that without it, does not depend on the delay time. 
\label{fig:pube}}
\end{figure}

The energy response in the 137~keV expected region for delayed signals was calibrated with
measurements collected in the presence of a $^{139}$Ce radioactive source (165.9~keV $\gamma$-line).
In the first successful test following the calibration measurements,  a highly
active ($>$10$^4 \, {\rm n} \,{\rm s}^{-1}$) PuBe neutron source was used
 to see if the number of neutron events could be measured using a
NaI scintillator via the suggested method (Fig.~\ref{fig:pube}). This test did not involve any special moderators: 
fast neutrons from the source were thermalized by surrounding materials.

Ambient neutron measurements using the aforementioned spectrometer were conducted  in one of the
buildings of JINR's Laboratory of Nuclear Problems (Dubna, Russia, $\sim$120~m above sea level).
Simultaneously, a low-background $^3$He neutron detector (CHM-57~\cite{chm57,rozov2010monitoring,rozov2010monitoring1})
also performed neutron flux measurements, as shown in Fig.~\ref{fig:nai-spectrometer}.
This $^3$He detector had a  neutron sensitivity, which refers here and hereafter to the isotropic, $4\pi$,
$1\, {\rm n}\, {\rm cm}^{-2}\, {\rm s}^{-1}$ thermal neutron flux in the detector's absence,
of $243\pm24\, {\rm counts}\, {\rm s}^{-1}$.

\begin{figure}
 \includegraphics[width=1.0\textwidth]{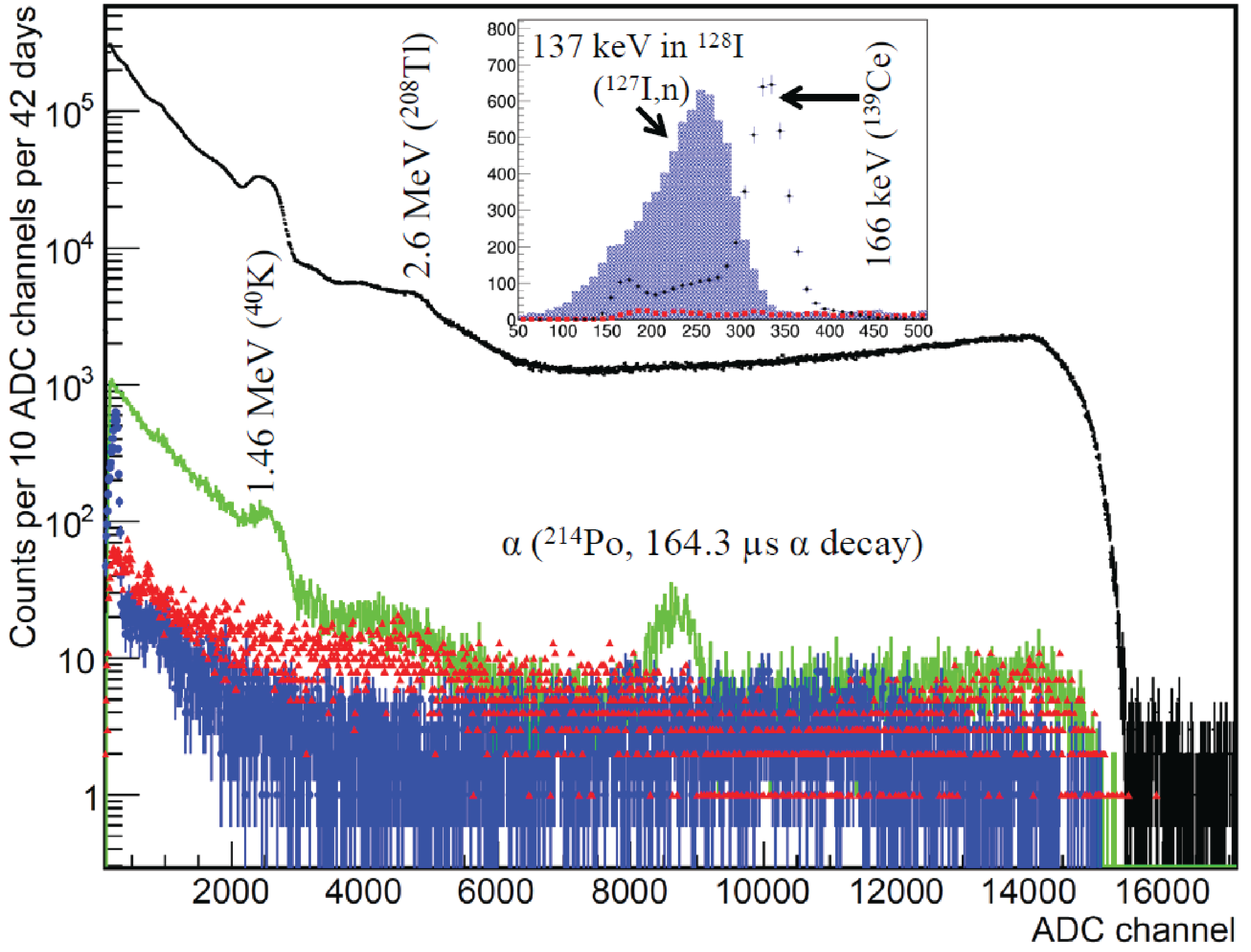}%
 \caption{The experimental energy spectra measured during 42~days in the LNP, 
JINR with a $^{63\times 63}$NaI spectrometer. 
Black dots - energy spectrum without any cuts; 
blue dots with error bars and red triangles - neutron prompt and neutron delayed spectra, 
respectively (delay time window is $1.8-10~\mu {\rm s}$, 
no ADC window cuts were applied, neutron events in the delayed spectrum are in a peak corresponding to ADC channels 150-350); 
green dots - delayed events with delay time window from 10 to 500~$\mu {\rm s}$. 
Insert: blue filled area - delayed event spectrum with delay time window 
from 1.8 to 10~$\mu {\rm s}$ (i.e. neutron events), asymmetrical peak shape caused by prompt signal influence on delayed signals; 
red squares - delayed event spectrum with shifted delay time window (11.8 to 20~$\mu {\rm s}$); dots with error bars - 
calibration with $^{139}$Ce 
(experimental calibration spectrum scaled down to match amplitude of delayed event spectrum).\label{fig:nai-spectra}}
\end{figure}

\begin{figure}
 \includegraphics[width=1.0\textwidth]{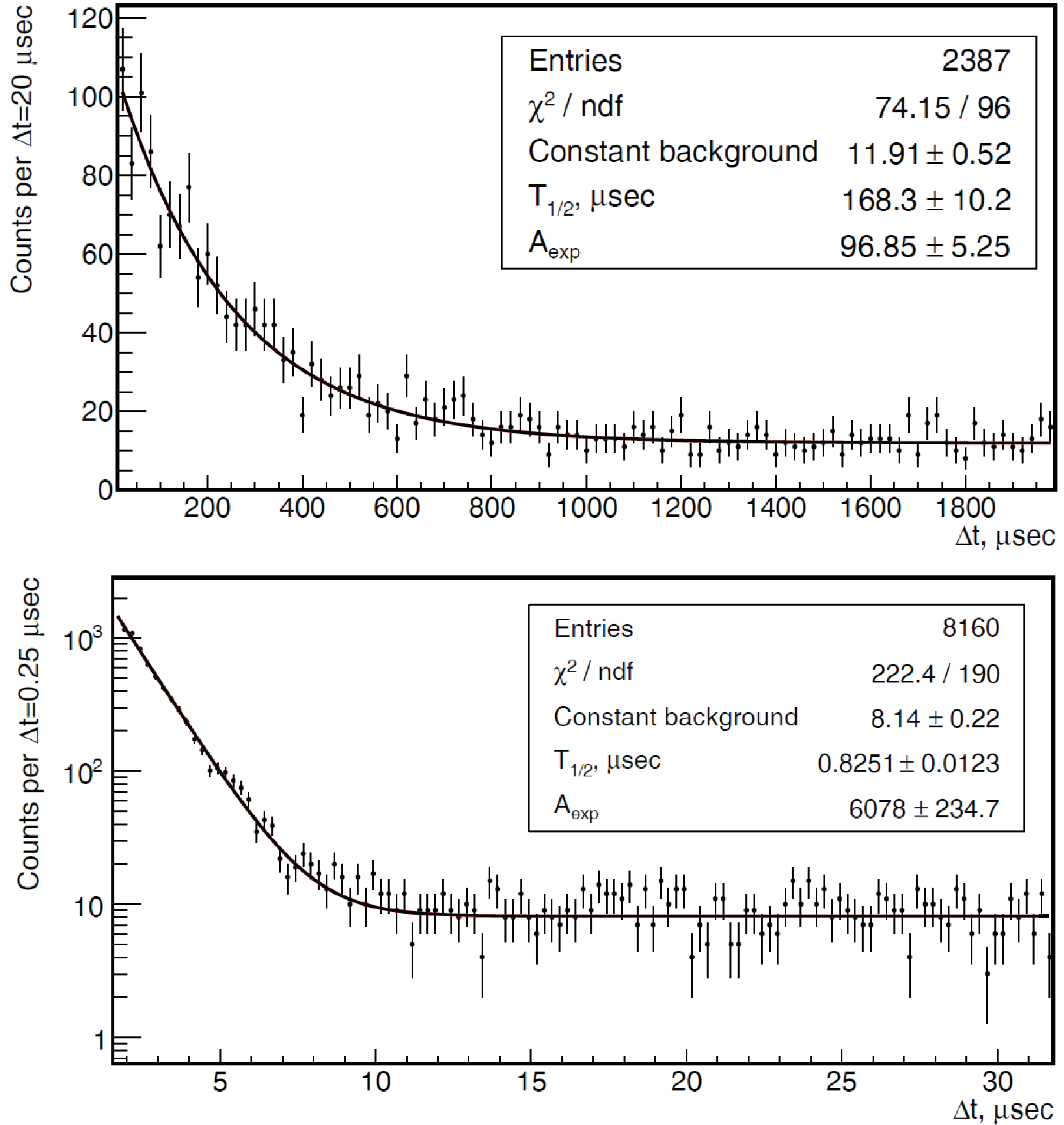}%
 \caption{Delayed coincidences. 
Upper graph shows $^{214}$Po's $\alpha$-decay. 
Lower decay curve illustrates 137~keV level's T$_{1/2}$. 
On both plots, dots with error bars correspond to observed coincidences with particular delay times. 
Solid lines show the fit of the function: $f=constant\_background+A_{exp}e^{ln(2) t/T_{1/2}}$, 
from which the half-lives of the data sets can be determined. 
\label{fig:nai-decay-time}}
\end{figure}

Fig.~\ref{fig:nai-spectra} displays the experimental energy spectra obtained by
the NaI(Tl) detector over the course of 42 days of measurements from May to June, 2016.
As evident from these spectra, it is impossible to discern neutron events from background signals without
using coincidence detection techniques to
distinguish the delayed event line in the 137~keV energy region.
The T$_{1/2}~{\rm of}~825\pm 12$~ns (Fig.~\ref{fig:nai-decay-time}),
obtained from measurements for this energy level, agrees with the tabulated value of $845\pm 20$~ns~\cite{tableofisotopes}.
Since systematic effects on the determination of T$_{1/2}$, such as event time precision, were not studied in this work, the acquired T$_{1/2}$ value 
should not be considered an improvement of the tabulated one.   
 The $^{214}$Bi$-^{214}$Po peak in coincidence events in data from measurements
with a delay time window of 10 to 500~$\mu$s helps to indirectly verify the delay time window approach to identifying coincidence events.
The experimentally acquired T$_{1/2}=168.3 \pm 10.8$~$\mu {\rm s}$  for $^{214}$Po agrees
with the tabulated value of $164.3\pm2.0~\mu {\rm s}$~\cite{tableofisotopes}.

\begin{figure}
\includegraphics[width=1.0\textwidth]{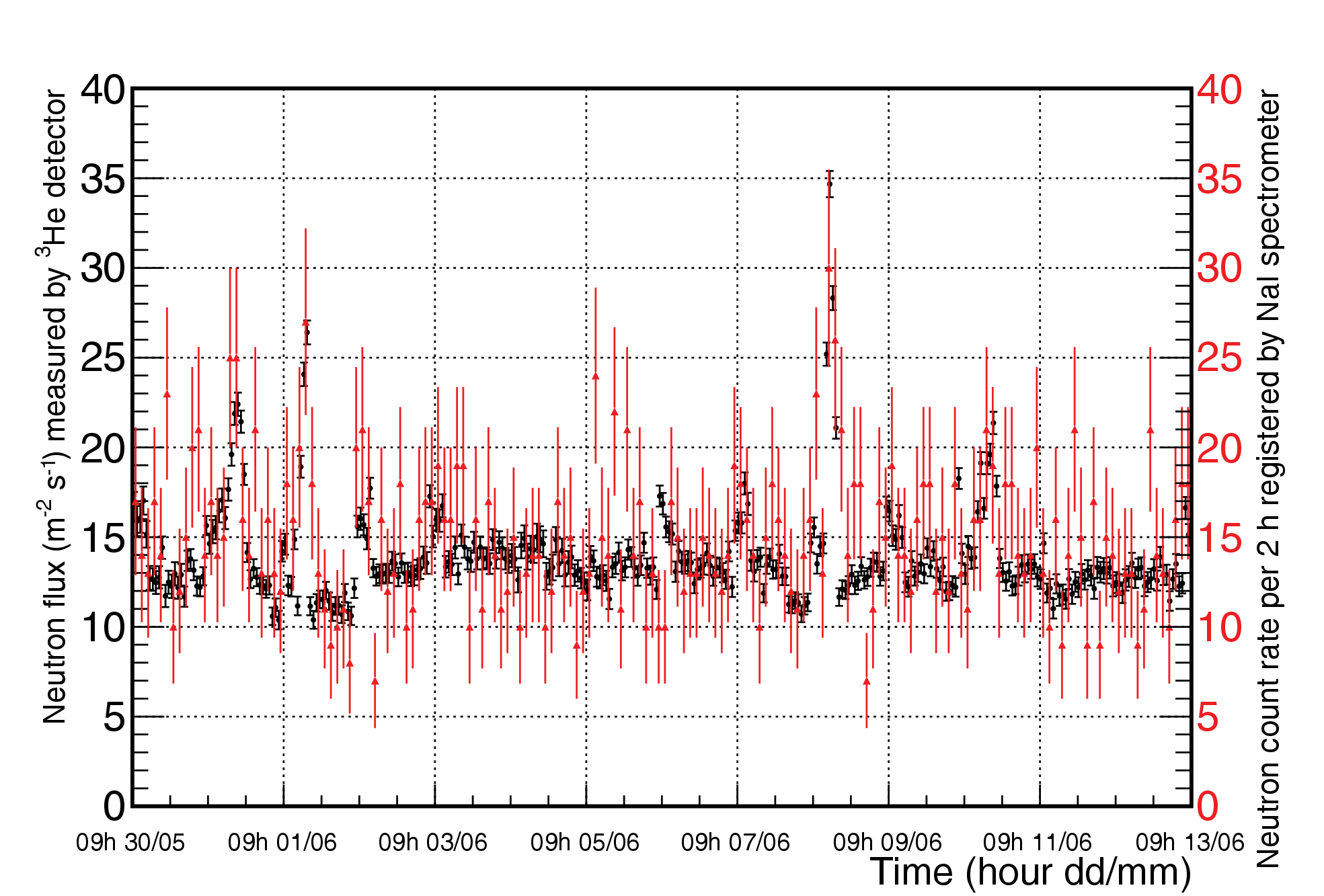}%
\caption{Comparison of neutron flux measured by $^3$He detector and number of neutron events registered
by NaI spectrometer (from measurements taken in May-June, 2016).
Black dots with error bars - neutron flux from $^3$He detector;
red triangles with error bars - neutron count rate from $^{63\times 63}$NaI spectrometer.\label{fig:nai-3he-comparison}}
\end{figure}

Fig. ~\ref{fig:nai-3he-comparison} compares the neutron flux
measured by the $^{63\times 63}$NaI detector with that measured simultaneously by the low-background
proportional $^3$He counter, thereby demonstrating the sensitivity of the NaI-based neutron detection approach.
Both spectrometers detected the small changes in ambient neutron
flux, on the level of $10^{-3}\, {\rm n}\, {\rm cm}^{-2}\, {\rm s}^{-1}$,
caused by the operation of one of JINR's acceleration facilities.
Based on analyses of the simultaneous measurements from the two detectors, the absolute sensitivity of
the $^{63\times 63}$NaI spectrometer was found to be $6.5 \pm 1 \, {\rm counts}\, {\rm s}^{-1}$ for a neutron flux of
$1\, {\rm n}\, {\rm cm}^{-2}\, {\rm s}^{-1}$ (without accounting for loss of effectiveness due to the 1.8~$\mu$s trigger holdoff).
Meanwhile, the number of coincidence events occurring within a delay time window of 1~$\mu {\rm s}$
was determined to be $0.8~{\rm day}^{-1}$ based on the constant background signal for a delay time 
greater than $10\, \mu {\rm s}$ (see Fig.~\ref{fig:nai-decay-time}), as well as on matching calculations with known
event count rates above the energy threshold and in the 137~keV energy region.

\section{Discussion}
Multiplying the event count rates in the prompt and delayed energy regions yields the frequency of coincidence events 
comprising the background signal for the suggested method of neutron measurement, 
which is almost quadratically proportional to the overall background signal.
Thus, the proposed method is especially sensitive in low-background experiments that use
iodine-containing detectors with masses of 10s or 100s of kilograms as active veto systems or as the main detectors.
Some examples of these types of experiments include the DAMA/LIBRA dark matter search experiment,
which uses a 250~kg NaI(Tl) detector~\cite{Bernabei2008297};
the KIMS dark matter search experiment, which implements a 103~kg CsI(Tl) detector~\cite{PhysRevD.90.052006};
the $\nu$GeN reactor neutrino nuclear coherent scattering experiment, which uses an approximately 400~kg NaI(Tl)
anti-Compton veto system~\cite{nuGeN};
and the COHERENT experiment, which uses a 14~kg CsI(Na) detector to search for coherent elastic neutrino scattering
from the Spallation Neutron Source ~\cite{Akimov:2015nza}.

In the aforementioned experiments,
neutrons comprise one of the most uncertain sources of background events.
Considering the sensitivity calculated for the $^{63\times 63}$NaI scintillator described in this work,
it is also possible to estimate the ability to detect the level and stability of thermal neutron flux in low-background
experiments. For a thermal neutron flux on the level of 10$^{-8}\, {\rm n}\, {\rm cm}^{-2}\, {\rm s}^{-1}$, 
approximately 1 neutron per day can be detected with a 100~kg NaI detector 
(the precise number can only be estimated after accounting for the influence of the detector's geometry on neutron capture efficiency 
and the detection of prompt and delayed signals).
For many low background experiments, only the detection of fast neutrons, which is a difficult and often even impossible task, is of interest.  
However, since no thermal neutron sources exist in nature, the thermal neutron flux is always in equilibrium with the fast neutron flux. 
Thus, despite the typically unknown ratio of fast to thermal neutrons, the measurement of thermal neutrons provides 
useful information about the total neutron flux and stability. For example, in~\cite{Bernabei2008297}, the upper limit on the thermal neutron
flux passing through the multicomponent DAMA/LIBRA shield was determined to be: $<1.2\, \times 10^{-7}\, {\rm cm}^{-2}\, {\rm s}^{-1}$~(90\% C.L.).
 
The majority of existing neutron detectors
are used for nuclear safeguards, security, and reactor instrumentation. 
The \textquotedblleft golden standard\textquotedblright ~among such detectors is the $^{3}$He-filled proportional counter with its high neutron
 detection efficiency and simultaneously low sensitivity to $\gamma$-radiation. 
A thorough comparison of NaI detectors with $^3$He detectors in the particular case of neutron portal monitors 
already exists in~\cite{ Holm201359}. Along with all other iodine-containing detectors, 
NaI spectrometers with comparable active volumes to $^3$He-based portal monitors have similar detection efficiencies 
and minimum detectable activities. The strong suppression of $\gamma$- background radiation 
marks the main advantage of the delayed coincidence approach to the detection of neutrons via an
 NaI (CsI) spectrometer. At the same time, the $1.8 \, \mu{\rm s}$ trigger holdoff and the fact that not all of the 
transitions pass through the delayed 137~keV level reduce the efficiency by a factor of $\sim$10, 
which can be prevented through improved data acquisition and/or a different light detector. 
To implement the proposed approach in a high intensity neutron field, the important, 
degrading effect of high-dose fast neutron bombardment on crystals must be taken into account.

For large detectors, such as that used in~\cite{Holm201359} or larger, a subsequent measurement of the total,
high-energy transition cascade (above the natural radioactivity threshold) coupled with measurements from the
delayed 137.8~keV energy level can effectively suppress background signals, even with a high $\gamma$-count rate.
This differs from the NaI detector used in this experiment and other relatively small detectors,
where only a fraction of events in the high-energy cascade from 6.8~MeV to 137.8~keV are registered.

Another important point of discussion is the 
detection of neutron events with a high multiplicity 
(neutrons from spontaneous fission, those produced by high energy muons in surrounding detector materials, etc). 
The proposed method can measure such events only after careful consideration and changes in the experiment: 
conducting the experiment with a short delay time window for prompt-delayed 
events of $<3 \, \mu{\rm s}$ would likely still not exceed the average time between neutron 
production and capture in the detector.

The decay process of $^{128}$I has another notable excited state at 167.4~keV,
which has a T$_{1/2}$ of $175\pm15$~ns (Table~\ref{tab:127i-scheme}).
This energy level is emptied exclusively via a 29.6~keV transition to the aforementioned 137.8~keV energy level,
providing a potential way to radically suppress background signals through triple coincidence measurements.
However, this goal necessitates the detection of delayed events
in a time window of 100s of ns. This is a difficult task when using NaI, and especially CsI, scintillators because
they have a decay time of $\sim$200 $-$ 500~ns. Triple coincidence measurements were not investigated in this research.

A similar neutron detection method can be implemented for Ge-detectors by using $^{73}$Ge's delayed
transitions from the 13.3~keV (T$_{1/2}=2.95\, \mu{\rm s}$) and 66.7~keV (T$_{1/2}=0.499\, {\rm s}$)
energy levels, where the latter is only helpful in ultra-low-background measurements.
However, it is important to note several experimental differences: $^{72}$Ge's thermal neutron capture cross section
of 0.98~b is 6 times smaller than that of $^{127}$I, and the $^{72}$Ge isotope only comprises 27.7\% of the
atoms in $^{\rm nat}$Ge. Furthermore, unlike $^{128}$I, $^{73}$Ge is a stable isotope,
so it can be excited to the aforementioned energy levels by fast neutron inelastic scattering,
causing ambiguous results.  For low-background neutron measurements, the cosmogenic
isotope $^{73}$As's decay needs to be accounted for in the analysis. Also, in a Ge-detector it is necessary to
consider the influence of the cold cryostat, the amount of nitrogen in the Dewar vessel, and the vessel itself on the
incoming neutron energy distribution (i.e. the neutron capture cross section).

\section{Conclusion}
In summary, this work proposes a fundamentally new neutron detection technique involving
spectrometers with existing, iodine-containing (NaI or CsI) scintillation detectors.
The additional use of delayed coincidence techniques allows for the highly effective discrimination
of neutron events from background signals. The precise measurements from a simple, inexpensive NaI(Tl) spectrometer
of the ambient neutron flux at sea-level demonstrate the sensitivity of this technique.

\section*{Acknowledgments}
This work has been funded by the Russian Science Foundation under grant 14-12-00920.
A.~Drokhlyansky would like to thank MIT Science and Technology Initiatives for their support.

\section*{References}
\bibliography{yakushev-nai}

\end{document}